\documentstyle[epsf,12pt]{article}
\newcommand{\lsim}{\mbox{\raisebox{-.3em}{$\stackrel{<}{\sim}$}}}

\renewcommand{\cite}[1]{\ref{#1}}
\newcommand{\half}{\frac{1}{2}}

\begin{document} 
\baselineskip=0.5cm
\begin{center}
{\bf How natural is a small but nonzero cosmological constant?}\vspace{1cm}\\
Yasunori Fujii\footnote{E-mail address: ysfujii@tansei.cc.u-tokyo.ac.jp}\\
Nihon Fukushi University,
Handa, Aichi, 475\ Japan\\
and\\
Institute for Cosmic Ray Research,
University of Tokyo, Tanashi, Tokyo, 188 Japan \vspace{.5cm}\\
\end{center}
\begin{abstract}
Based on our previous attempt, we propose a better way to understand a small but nonzero cosmological constant, as indicated by a number of recent observational studies.  
We re-examine the assumptions of our model of two scalar fields, trying to explain the basic mechanism resulting in a series of mini-inflations occuring nearly periodically with respect to $\ln t$ with $t$ the cosmic time.  
We also discuss how likely the solution of this type would be, depending on the choice of the parameters. 
\end{abstract}
\vspace{.5cm}A growing number of different  observations, notably the recent determination of the Hubble constant $H_{0}$ [\cite{fhp}], seem to point to a suggestion that there is a possible {\em small but nonzero cosmological constant} $\Lambda$ [\cite{os}], with $\Omega_{\Lambda}\equiv\Lambda /\rho_{\rm cr} \lsim 1$, where $ \rho_{\rm cr}=(3/8\pi G)H^{2}_{0}$.
  This may not, however, be readily acceptable from a theoretical point of view, because introducing $\Lambda$  has been considered to be highly {\it ad hoc}.  
Contrary to this long-held prejudice, on the other hand, it is widely recognized that a cosmological constant is an indispensable ingredient in many of the theoretical models of unification.\footnote{Some authors [\cite{fkyn}] assume $\Lambda =0$ in the starting Lagrangian.  This is protected, however, by no known symmetries against any likely perturbations.} 
Unfortunately, they tend to predict  $\Lambda$ larger than the observed value, or its upper bound, by as much as 120 or so orders of magnitude.  
One of the possible ways out is to devise a theory in which the cosmological constant is not a true constant but decays like $\sim t^{-2}$, with $t$ the cosmic time [\cite{dol},\cite{yf1}].

Notice that the theoretically natural size of $\Lambda$ is of the order one in the Planckian unit system with $8\pi G=1$,\footnote {By also choosing $\hbar = c=1$, the units of length, time and the energy are 8.09$\times 10^{-33}$cm, 
2.70$\times 10^{-43}$sec, and 2.44$\times 10^{18}$GeV, respectively.} while the present age of the Universe $t_{0}\sim 10^{10}$y being of the order of $10^{60}$.  
In this scenario of ``a decaying cosmological constant," {\em today's cosmological ``constant" is small $\lsim 10^{-120}$ only because our Universe is old.}  No unnaturally extreme fine-tuning of  parameters is called for.

The scenario  has been often formulated based on the models in which a scalar field plays a role; the time-dependent {\em effective} cosmological constant 
  $\Lambda_{\rm eff}(t)$ is in fact the energy density of this scalar field that couples to the ordinary matter only as weakly as gravity.

This model implies, however, a complete {\em  absence} of the cosmological {\em constant}.  What is indicated by the observation is, on the contrary, the {\em presence} of a flat portion in the energy density as a function of $t$, a deviation from a smooth falling-off $\sim t^{-2}$.   
We came across, however, to a model which, by introducing another scalar field, would result in occasional flattenings of $\Lambda_{\rm eff}(t)$ thanks to  a nonlinear nature  of the cosmological equations [\cite{pl}].  
The purpose of this note is to provide a simpler understanding of the mechanism.

It seems appropriate here to state our attitude.  
In view of the lack of the
  complete theoretical framework to derive all the details of the final
 results, we follow a heuristic approach, trying to see what the effective theory in 4
 dimensions should be like, if it is to fit to what appears to be the  effect of a  small but nonzero
 cosmological constant. As it turns out, this is highly nontrivial, if the model is somehow related to modern unification theories.  We list some  of the main assumptions in Ref. [\cite{pl}].

First we assume the presence of a scalar field $\phi$ of the
 {\it dilaton}-type, having a non-minimal coupling, which is chosen,
 for the sake of simplicity, to be a Brans-Dicke type; $\phi^2 R$.\footnote{Our $\phi$ is, following the standard notation in the conventional relativistic field theory, related to the original notation $\varphi$ in Ref. [\cite{bd}] by 
$\varphi= \phi^{2}/8\omega$.}  
We then
 apply a conformal transformation (Weyl rescaling) to remove the non-minimal
 coupling.  
We do this for the technical convenience, at the moment, though
 the correct  conformal frame (CF) should be selected according to what clock we use
 to describe the evolution of the Universe.  In this connection we should
 notice that none of the realistic theories of gravity is conformally invariant, and that our conclusion  remains true also in the
 original CF in which the non-minimal coupling is present.

As an important consequence of this transformation the $\Lambda$ term in the original CF is converted to a potential of $\phi$ of the type $\Lambda e^{-\sigma /\kappa}$ where $\kappa$ is a constant\footnote{See Ref. [\cite{yf1}] on how $\kappa$ is constrained in order for the results to be consistent with realistic cosmology.} while $\sigma$ is a transformed scalar field appropriate in the new CF.  
The $\sigma$ field rolls down the slope toward infinity, ensuring the $\Lambda_{\rm eff}$,  essentially the energy of $\sigma$, to fall off like $\sim t^{-2}$ after the inflationary era.

As a next step we introduce another scalar field $\Phi$ which has a specific interaction with $\sigma$ but couples to conventional matter fields also as weakly as the gravitational interaction.  
We discovered an example of the 
interaction such that $\Lambda_{\rm eff}$, which is now the total energy density of the $\sigma\:$-$\Phi$ system, shows a repeated occurrence of leveling-off 
superimposed on the overall smooth fall-off $\sim t^{-2}$, coming barely to exceed the normal matter density, hence acting as a cosmological constant.  In accordance with this the scale factor $a(t)$ exhibits an extra acceleration deviating from the overall smooth behavior $\sim t^{1/2}$ or $\sim t^{2/3}$.  
A typical solution of the cosmological equations was shown in Fig. 3 of Ref. [\cite{pl}].

The basic cosmological equations are ($k=0$) [\cite{pl}]
\begin{eqnarray}
&&3H^2 = \rho_{\rm s}+\rho_{\rm m}, \label{cc2_1}\\
\nonumber\\
&&\ddot{\sigma}+3H\dot{\sigma}- \exp (-\sigma /\kappa)
  \left[ \frac{\Lambda}{\kappa} +\half m^2 \Phi^2 \left( 
    \frac{U}{\kappa}-\frac{dU}{d\sigma} \right)   \right] =0,\label{cc2_2}\\
\nonumber\\
&&\ddot{\Phi}+3H\dot{\Phi}+ \exp (-\sigma /\kappa)m^2 U\Phi=0,\label{cc2_3}
\end{eqnarray}
where
\begin{equation}
\rho_{\rm s}=\half \dot{\sigma}^2 +\half \dot{\Phi}^2 + V,
\label{cc2_4}
\end{equation}
and
\begin{equation}
V(\Phi,\sigma) =\exp (-\sigma /\kappa) 
\left[ \Lambda +\half m^2 \Phi^2 U(\sigma)\right],
\label{cc2_5}
\end{equation}
with
\begin{equation}
U(\sigma)= 1+B\sin(\omega \sigma).
\label{cc2_6}
\end{equation}
Here $\kappa, m, B$ and $\omega$ are constants.  
The exponential factor 
$\exp(-\sigma /\kappa)\;(\sim \phi^{-4})$ comes typically from the Weyl rescaling, thus transforming the $\Lambda$-term into the potential, as mentioned before.  
The sinusoidal dependence in (\ref{cc2_6}), on the other hand, has been introduced on the try-and-error basis.  
We later discuss how this specific form is favored.  
It is crucial to assume that the conformal transformation property of $\Phi$ is such that the same factor $\exp (-\sigma /\kappa) $ appears in front of the $\Phi^2 U(\sigma)$ term.

For the matter density $\rho_{\rm m}$ we assume the mixture of relativistic and non-relativistic matters:
\begin{equation}
\rho_{\rm m}=\rho_{\rm r}a^{-4}+\rho_{\rm nr}a^{-3}.
\label{cc2_6a}
\end{equation}

We admit that the result depends heavily on these assumptions.  
In view of the huge discrepancy of 120 orders of magnitude between conventional theory and the observation, however, variety of models seem to deserve consideration as working hypotheses.  
Notice that all of our parameters are essentially of the order one in  Planckian units, appealing to {\em theoretical naturalness}.

One might still argue that we are introducing as much as what we want to come up with.  
We emphasize, however, that it is far from trivial to make a right choice on what to be introduced; otherwise it will not make sense no matter how much we bring in.

On using the new time variable $\tau$ defined by
\[
\tau\equiv \ln t, 
\]
and also defining $b(\tau) $ by
\[
a= e^{b},
\]
eqs. (\ref{cc2_1})-(\ref{cc2_3}) are put into
\begin{eqnarray}
&&3 b'\:^2 =\half \sigma'\:^2 +\half \Phi'\:^2 + t^2 \left(V +\rho_{\rm m}\right), \label{cc2_7}\\
\nonumber\\
&&\sigma'' +(3b' -1)\sigma' -t^2\exp (-\sigma /\kappa)\left[ \frac{\Lambda}{\kappa} +\half m^2 \Phi^2 \left( 
    \frac{U}{\kappa}-\frac{dU}{d\sigma} \right)   \right]=0,\label{cc2_8}\\
\nonumber\\
&&\Phi''+(3b' -1)\Phi'+ t^2\exp (-\sigma /\kappa)m^2 U\Phi=0,\label{cc2_9}
\end{eqnarray}
where $'$ means differentiation with respect to $\tau$.  Notice the explicit occurrence of the time variable $t^{2}$ on the right-hand sides.

It is also worth noticing that (\ref{cc2_7})-(\ref{cc2_9}) allow the asymptotic solution for $t\rightarrow\infty$ if $U$ is chosen to be constant, namely $B=0$:
\begin{eqnarray}
a(t)&=& t^{2/3},\quad\mbox{or}\quad b(\tau)=\frac{2}{3}\tau,\label{cc2_10}\\
\nonumber\\
\sigma(t)&=&2\kappa \ln\left( \sqrt{\frac{\Lambda}{2}}\frac{t}{\kappa}  \right),\label{cc2_11}\\
\nonumber\\
\Phi(\tau)&=&A e^{-\tau /2}\sin (\tilde{m}\tau),
\label{cc2_12}
\end{eqnarray}
where $A$ is an integration constant while
\[
\tilde{m}=\sqrt{\frac{2\kappa^2 m^2}{\Lambda}-\frac{1}{4}}.
\]

We learn that the $\Phi$ field, if decoupled from $\sigma$, would play no role in the asymptotic era and that   $\tau$ might be a useful variable.  
From the solution (\ref{cc2_11}) also follows that the combination 
\[
F(t, \sigma)\equiv t^2\exp(-\sigma /\kappa) 
\]
tends to a constant ($=2\kappa^2 /\Lambda$) if $U=1$.  This implies that with non-constant $U(\sigma)$ this combination might be nontrivial.  We show that this is indeed the case.

A typical solution obtained numerically is shown in Fig. 1, another example  with  parameters somewhat different from those used in Fig. 3 in Ref. [\cite{pl}].  
Notice that we chose the ``initial time" $t_{1}=10^{10}$, because, though the real classical cosmology had begun much earlier, we can conveniently avoid more details on the inflation era and the ensuing reheating process.

We first find in the plot (a), as already alluded, the scale factor $a(t)$ shows a series of ``mini-inflations,'' each implying a rapid increase lasting during an opoch which is ``short" in terms of $\tau$, but could be quite ``long" if it is measured in the ordinary time $t$; nearly as comparable as $t$ itself.  
One of such epochs has been chosen to include the present time with $\lambda \equiv \log t\approx$60, 
corresponding to $t \approx 10^{10}$y. (The scale factor resumes a usual expansion immediately beyond the frame.)

In the plot (b) we notice that $\rho_{\rm s}$, the total energy of the $\sigma$-$\Phi$ system, and $\rho_{\rm m}$, the ordinary matter energy density, fall off like $\sim t^{-2}$ as an over-all behavior, but interwinding each other.  
A closer look reveals that a mini-inflation occurs whenever $\rho_{\rm s}$ exceeds $\rho_{\rm m}$.

Fig. 2 is the same plot as Fig. 1 (b) presented in a magnified scale 
around the present time; showing that $\rho_{\rm s}$ surpasses $\rho_{\rm m}$, remaining nearly constant for a while, hence {\em imitating a cosmological constant with the size basically of the same order of magnitude as} $t^{-2}\sim \rho_{\rm cr}$.

We also notice in Fig. 1(a) that each of these anomalous behaviors takes place toward the end of the ``dormant period,'' during which both of the scalar fields come almost to standstill.  To understand this behavior we first point out that the dormant period and its repetition are primarily due to the dynamics of the  $\sigma$-$\Phi$ system; the ``back-reaction'' from the cosmological expansion has a rather minor effect.

Fig. 3 shows an example in which the cosmological part is cut off with the same parameters but with $3b' -1$ replaced  by 0.5.
In spite of some differences, which, representing how much the cosmological effect could be, will be discussed shortly, 
comparing Fig. 1(a) and Fig. 3 is sufficient to  convince ourselves that  the ``recycled dormant periods" could take place even without cosmological effect. 
From this point of view, we now focus upon more detailed analysis of the solution in the {\em isolated} $\sigma$ -$\Phi$ system as a simplified mathematical model.

The initial value $\sigma_{1}=6.75442$ implies that $\omega\sigma_{1}= 2\pi\times 10.75$; the $\sigma$ field starts at one of the potential minima as given by $\sin (\omega\sigma)$ if viewed in the $\sigma$ direction (see Fig. 2 of Ref. [\cite{pl}]), but on the slope in the direction of $\Phi$.  
Also the ``initial time'' chosen to be $t_{1}=10^{10}$ implies $2\tau_{1}=46.052$, and hence $F(t_{1}, \sigma_{1})= t_{1}^2\exp(-\sigma_{1} /\kappa)=\exp(2\tau_{1} -\sigma_{1}/\kappa )=\exp(3.303)=27.19$, which is quite large.  
As a result, $\sigma$ is pushed forward strongly.  In this sense the system started from a ``catapulting stage."  
The rapid increase of $\sigma$, however, makes $F(t,\sigma)$ small, as will be found by comparing the curve of $\sigma$ and the straight line $2\kappa\tau$, also shown in Fig. 3.  Soon $\sigma$ is nearly free, going further until it is decelerated by the frictional term $3b'-1\approx 0.5$, finally to be trapped to another minimum of $\sin (\omega\sigma)$.

On the other hand, the  $\Phi$ field, having pearched on the middle of the potential slope, is also catapulted downward, passing the central valley $\Phi =0$ past, until the force dwindling again due to the decrease of $F(t, \sigma)$ and the cosmological friction stop it to an almost complete halt, hence the dormant period.    
The energy density $\rho_{\rm s}$ still continues to decrease according to $\sim t^{-3}$ before it stays constant.\footnote{See also Ref. [\cite{om}] for a similar behavior.}

Now with a virtually unchanging $\sigma$, the increasing $\tau$ makes $F(t, \sigma)$  non-negligible again, bringing the system back to the catapulting stage from which we started before.  
In this way the dormant period may repeat itself nearly periodically with respect to $\tau$ (instead of  $t$), if the field configurations match suffciently close to the previous values.

The ``recycling," however, may fail if the matching turns out incomplete.  
Fig. 3 is in fact one of the patterns of such ``short" recycling encountered most commonly.
Toward the end of a dormant period, $\sigma$ is ``released" off the track before it is kicked hard by the force which has been building up.
The system enters into an asymptotic behavior in which $\sigma$ grows linearly while $\Phi$ decreases slowly toward $\Phi =0$.  If the same behavior occurs in the cosmological setting, the scale factor increases smoothly, resulting in no effective cosmological constant today.

On the other hand, much ``longer" recycling is also rather  common, as shown in Fig. 4.  
Suppose a recycling process lasts suffciently long in the cosmological system.
Then, as in the isolated $\sigma$-$\Phi$ system, $\rho_{\rm s}$ would stay nearly constant toward the end of the dormant period, surpassing $\rho_{\rm m}$, hence playing the role of a cosmological constant.
When the scalar fields start moving as the factor $F(t, \sigma)$ increases, however, $\rho_{\rm s}$ begins to decrease, eventually nosediving beneath $\rho_{\rm m}$.  The Universe resumes an ordinary expansion again.  This explains the behaviors shown in Figs. 1-2.

A question then arises how likely the solutions of sufficiently long recycling could be.  An idea on the answer may be obtained again by studying the isolated $\sigma$-$\Phi$ model.  We surveyed solutions of the isolated $\sigma$-$\Phi$ system by changing one of the initial values, $\Phi_{1}$ for example, keeping other parameters and initial values fixed.\footnote{Presentation of the result has been made simpler by choosing parameters different from those used in Figs. 1-3, but the same as in Fig. 4 except for $\Phi_{1}$.}  
Solutions with shorter recycling exhibit basically the same patterns as in Fig. 3, while Fig. 4 is an example of sufficiently long recycling.
Combining the solutions, we plotted in Fig. 5, the time of the end of recycling, $t_{\rm e}$, against $\Phi_{1}$ varied discretely.
In spite of apparently rampant variation, we obtain solutions of long recycling ($\lambda_{\rm e}> 62$) for 8 out of 23 choices of $\Phi_{1}$.
This, together with other limited but similar examples, seems to be an encouraging sign that the occurrence of the continued recycling is reasonably likely.  
The same ``optimistic" view applies also if the cosmological effects are fully included.

We emphasize that the presence of minima of the potential as a funciton of $\sigma$ is crucial.  
The leveling-off behavior is triggered by trapping $\sigma$.  
The form $\sin(\omega\sigma )$ is favored because it is ready to trap $\sigma$ virtually at any time.  
Any other potential will probably be acceptable, from a ``phenomenological" point of view, if it shares this property.

We have shown that a small but nonzero cosmological constant as required by the observations could result rather 
naturally due to a nonlinear nature of the scalar field equations.  
As a favored coupling, we tentativley suggested a Sine-Gordon-like interaction, though its real origin is yet to be discussed [\cite{pl}]. 
Furthermore having introduced two scalar fields, we have too many parameters, including the initial conditions, to allow unique predictions, or even a systematic survey of the solutions.  
In this sense our conclusion is still preliminary.  
The example in Figs. 1-2 merely illustrates how the results {\em can} be realistic, leaving many details yet to be worked out.
We nevertheless have promising indications that the desired result comes about quite likely.

We certainly have to adjust some of the parameters in order to bring a mini-infaltion to the epoch including the present time, for example. The extent to which we are supposed to fine-tune them is rather {\em mild}, however.\footnote{Changing $\Phi_{1}$ from 0.212 to 0.210 in the solution of Figs. 1-2 would shift $\Omega_{\Lambda}=0.67$ at $\lambda =60.15$ ($t=1.21\times 10^{10}$y) to 0.71.  However, $\Phi_{1}=0.200$ gives $\Omega_{\Lambda}\sim 1-10^{-20}$.  On the other hand, $\Phi_{1}=0.2115$ yields $\Omega_{\Lambda}=0.73, H_{0}=$79 km/sec/Mpc at $t=1.5\times 10^{10}$y.}  
In other words, detailed analyses of various cosmological parameters at present, like $H_{0}, t_{0}, \Omega_{\Lambda}$ and $q_{0}$, will serve to constrain the parameters of the theory, as will be discussed elsewhere.

As one of the generic consequences of the present mechanism, we anticipate some past epochs to have emerged with significant amount of $\Omega_{\Lambda}$.  
In the solution of Figs. 1-2, for example, we find $\Omega_{\Lambda} \gg 1$ for $27\lsim \lambda \lsim 39$, while we have avoided the same at the epoch of primordial nucleo-synthesis ($\lambda\sim 45$).  
This may illustrate how the parameters and the initial conditions can be determined, in principle, also by looking into details of the past cosmological histories, which should deserve future studies.

We confined ourselves to the ``primordial" cosmological constant prepared in the starting Lagrangian.  It is yet to be seen if the same mechanism applies successfully to the vacuum energies associated with cosmological phase transitions at later times.

\begin{center}
{\large\bf References}
\end{center}
\begin{enumerate}
\item\label{fhp}W. Freedman {\it et al.}, Nature {\bf 371}, 757 (1994).  See also,  M. Fukugita, C.J. Hogan and P.J.E. Peebles, Nature, {\bf 366}, 309 (1993).
\item\label{os}See, for example, J.P. Ostriker and P.J. Steinhardt, Cosmic Concordance, and papers cited therrein.
\item\label{fkyn}See, for exmaple, M. Fukugita and T. Yanagida, Model for the Cosmological Constant, YITP/K-1098 (1994). 
\item\label{dol}A.D. Dolgov, in {\it The Very Early Universe}, Proceedings of Nuffield Workshop, England, 1982, edited by G.W. Gibbons and S.T. Siklos (Cambridge University Press, Cambridge, England, 1982): L.H. Ford, Phys. Rev. {\bf D35}, 2339(1987).
\item\label{yf1}Y. Fujii and T. Nishioka, Phys. Rev. {\bf D42}, 361(1990).
\item\label{pl}Y. Fujii and T. Nishioka, Phys. Lett. {\bf B254}, 347(1991).
\item\label{bd}C. Brans and R.H. Dicke, Phys. Rev. {\bf 124}, 925(1961).
\item\label{om}Y. Fujii, M. Omote and T. Nishioka, Prog. Theor. Phys. {\bf 92}, 521(1994).

\end{enumerate}
\newpage
\begin{figure}[h]
\epsffile{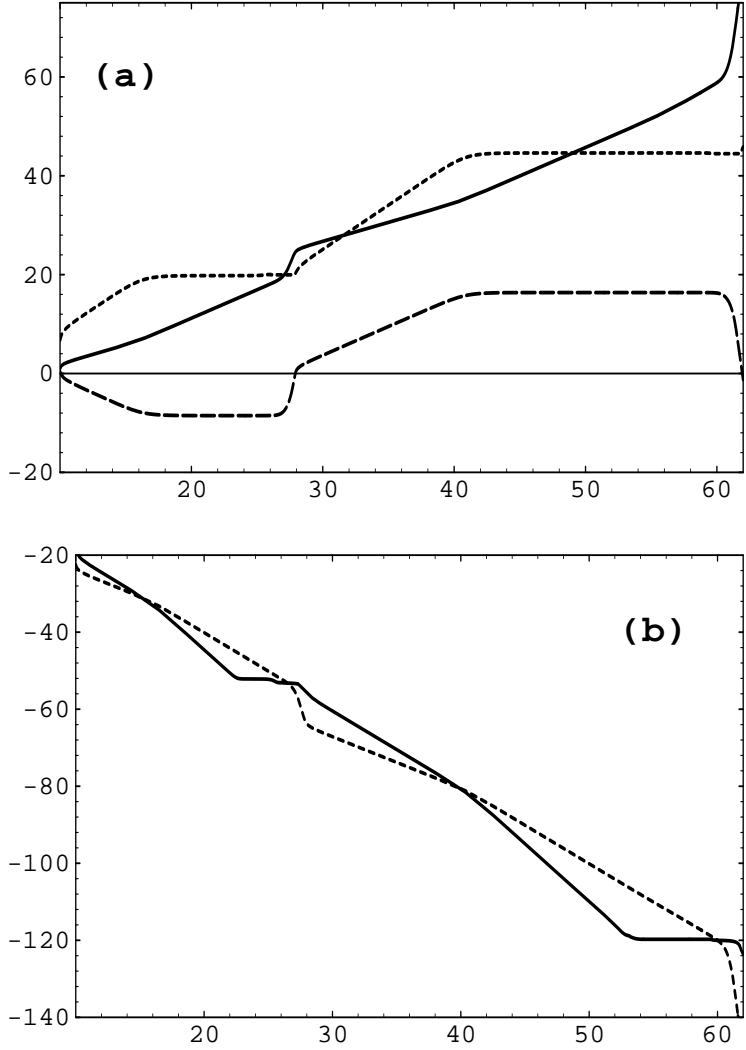}
\caption{An example of the solution of (8)-(10). (a) Upper plot: $b=\ln a$ (solid), $\sigma$ (dotted) and $2 \Phi$ (broken) are plotted against $\lambda =\log t= 0.434 \tau$. 
The present age of the Universe supposed to be (1.0\,-\,1.5)$\times 10^{10}$y corresponds to 60.0$\,$-$\,$60.2 of $\lambda$ in units of the Planck time.  
The parameters were chosen to be $\Lambda =1, \kappa =0.158, m=4.75, B=0.8, \omega =10$ in the Planckian units.  
The initial values chosen conveniently at $t_{1}=10^{10}$ are $a=1, \sigma_{1}=6.75442, \dot{\sigma}_{1}=0, \Phi_{1}=0.212, \dot{\Phi}_{1}=0, \rho_{\rm r 1}=2.04\times 10^{-21}, \rho_{\rm nr 1}=4.46\times 10^{-44}$; the last two being adjusted to give the ``equal time" $\lambda_{\rm eq}\sim 55$.
The value of $\sigma_{1}$ corresponds to starting at a minimum of $\sin(\omega\sigma)$.  (b) Lower plot: $\rho_{\rm s}$ (solid), the total energy density of $\sigma$ and $\Phi$, and $\rho_{\rm m}$ (dotted), the matter energy density, against $\lambda =\log t$.  
}
\end{figure}

\begin{figure}[h]
\epsffile{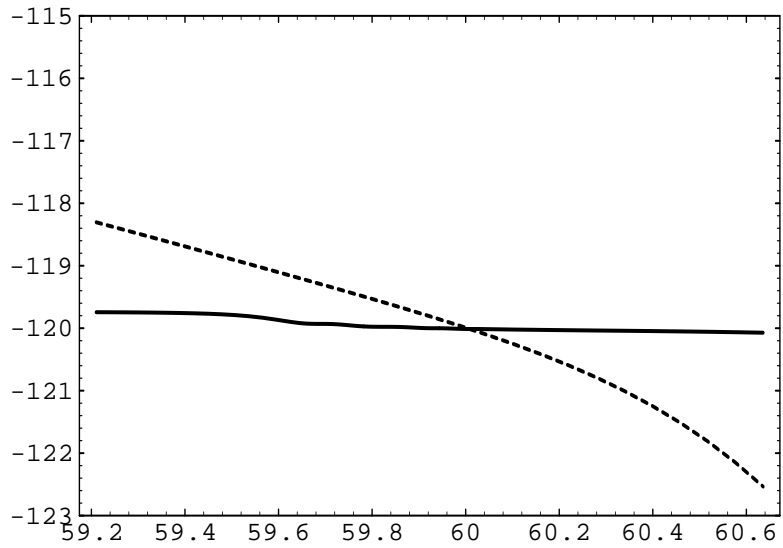}
\caption{The same plot as in Fig. 1(b) but in a magnified scale of $\lambda$ around the present time.
We find $\Omega_{\Lambda}=0.67$ and $H_{0}=81$km/sec/Mpc at $\lambda =60.15$ ($t=1.21\times 10^{10}$y).
}
\end{figure}

\begin{figure}[h]
\epsffile{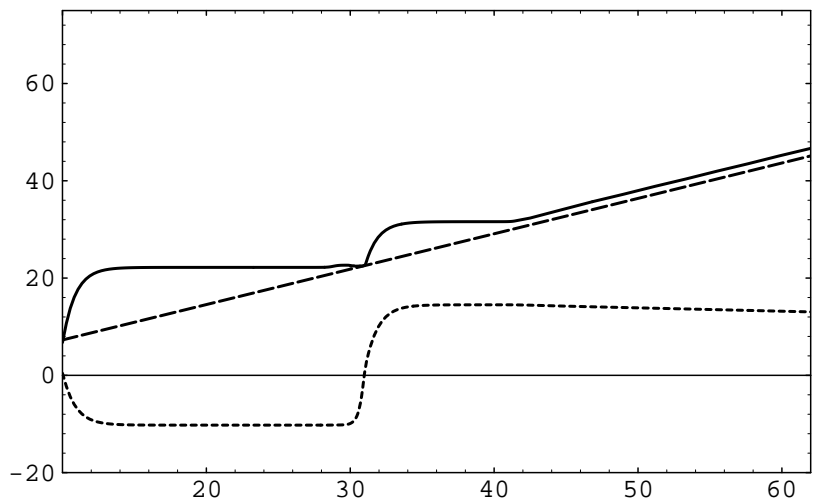}
\caption{An example of the solution in the isolated $\sigma$-$\Phi$ system, in which recycling of the dormant periods ends prematually at $\lambda\approx 41$. 
$\sigma$ (solid) and $2\Phi$ (dotted) are shown against $\lambda =\log t$.  Also shown is $2\kappa\tau$ (broken) to be compared with $\sigma$.
All the parameters remain the same as in Figs. 1-2, except for $3b'-1$ replaced by 0.5.  
}
\end{figure}

\begin{figure}[h]
\epsffile{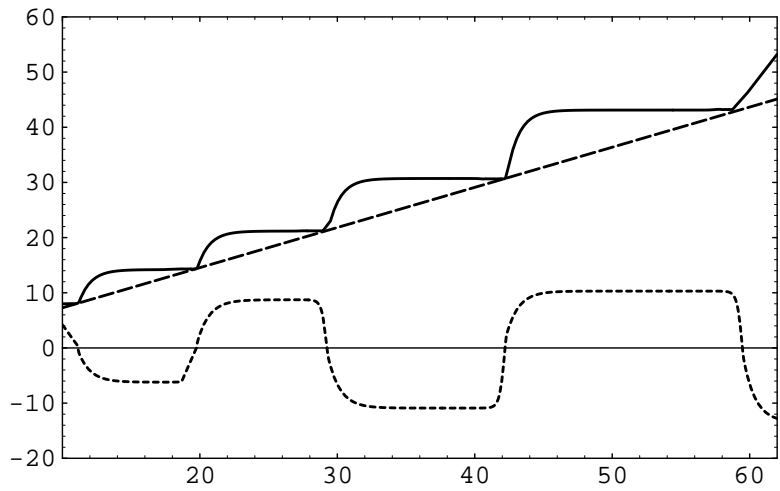}
\caption{An example of the solution in the isolated $\sigma$-$\Phi$ system, showing long recycling.  
We choose $m=5.0$ with other parameters as well as the symbols the same as in Fig. 3. 
The initial values at $\lambda_{1}=10$ are $\sigma_{1}=8.0, \dot{\sigma}_{1}=0, \Phi_{1}=2.1, \dot{\Phi}_{1}=0.19$, somewhat different from those in Fig. 3.
}
\mbox{}
\mbox{}
\end{figure}
\newpage
\begin{figure}[h]
\epsffile{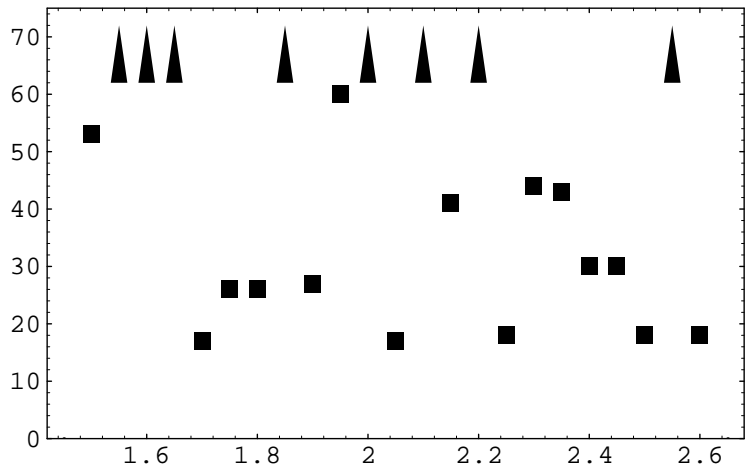}
\caption{The time $\lambda_{\rm e}$ for the end of recycling  in the isolated $\sigma$-$\Phi$ system is plotted against one of the initial values $\Phi_{1}$, varied with spacing 0.05.  The other initial values and the parameters are the same as in Fig. 4.  The arrows indicate $\lambda_{\rm e}> 62$.
}
\end{figure}

\end{document}